\documentclass[12pt,a4paper]{article}

\usepackage{graphicx}
\usepackage{amsmath}
\usepackage{amssymb}
\usepackage{epsfig,latexsym}
\newcommand{\be}{\begin{eqnarray}}
\newcommand{\ee}{\end{eqnarray}}

\begin{document}

\begin{centering}


{\bf \Large Constraints on cosmological birefringence from {\sc Planck} and Bicep2/Keck data}

\vskip 1 truecm

{\small {\sc A.~Gruppuso${}^{\; a,b}$,  M.~Gerbino${}^{\; c}$, P.~Natoli${}^{\; d}$, L.~Pagano${}^{\; c}$, \\ N.~Mandolesi${}^{\; d,a}$ and D.~Molinari${}^{\; d}$}} \\[20 pt]

{${}^a$\sl\small INAF-IASF Bologna, \\
Istituto di Astrofisica Spaziale e Fisica Cosmica di Bologna, \\
Istituto Nazionale di Astrofisica,
via Gobetti 101, I-40129 Bologna, Italy} \\
\vspace{8pt}

{${}^b$\sl\small INFN, Sezione di Bologna,
Via Irnerio 46, I-40126 Bologna, Italy} \\
\vspace{8pt}

{${}^c$\sl\small Dipartimento di Fisica and INFN, Universit\`a di Roma ÒLa SapienzaÓ, \\ 
P.le Aldo Moro 2, 00185, Rome, Italy} \\
\vspace{8pt}

{${}^d$\sl\small Dipartimento di Fisica e Scienze della Terra and INFN,
Universit\`a degli Studi di Ferrara, Via Saragat 1, I-44100 Ferrara, Italy}



\end{centering}

\date{\today}

\abstract{The polarization of cosmic microwave background (CMB)  can be used to constrain cosmological birefringence, the rotation 
of the linear polarization of CMB photons potentially induced by parity violating physics beyond the standard model. This effect produces non-null CMB cross correlations between temperature and B mode-polarization, and between E- and B-mode polarization. 
Both cross-correlations are otherwise null in the standard cosmological model.
We use the recently released 2015 {\sc Planck} likelihood 
in combination with the Bicep2/Keck/Planck (BKP) likelihood to constrain the birefringence angle $\alpha$. 
Our findings, that are compatible with no detection, read $\alpha = 0.0^{\circ} \pm 1.3^{\circ} \mbox{ (stat)} \pm 1^{\circ}  \mbox{ (sys)} $ for {\sc Planck} data and 
$\alpha = 0.30^{\circ} \pm 0.27^{\circ} \mbox{ (stat)} \pm 1^{\circ}  \mbox{(sys)} $ for BKP data.

We finally forecast the expected improvements over present constraints when the {\sc Planck} BB, TB and EB spectra at high $\ell$ will be included
in the analysis. }


\section{Introduction}

It has become customary to use polarized cosmic microwave background (CMB) data to constrain the cosmic birefringence effect \cite{Hinshaw:2013,Wu:2008qb,Kaufman:2013vbd}, 
i.e. the in vacuo rotation of the photon polarization direction during propagation \cite{Carroll:1989vb}.
In general, such effect, which is naturally parameterized by an angle $\alpha$,
results in a mixing between Q and U Stokes parameters that produces non-null CMB cross correlations between temperature and B mode-polarization, 
and between E- and B-mode polarization. 
Since these correlations are expected to be null under the parity conserving assumptions that is beneath the standard cosmological model, cosmic birefringence
is a tracer of parity violating physics. 

Several theoretical models exhibit cosmological birefringence. 
Any coupling of the electromagnetic (EM) field with axion-like particles \cite{Finelli:2008jv} or a quintessence field \cite{Giovannini:2004pf}
is expected to induce the effect, which can be also driven by quantum-gravity terms \cite{Gubitosi:2009} or Chern-Simons type interactions \cite{Carroll:1989vb} in the EM Lagrangian.
Each particular model predicts a specific dependence of the CMB angular power spectra (henceforth APS) on $\alpha$, which can be complicated.
In this paper, we restrict to the simplest case of constant $\alpha$, for which the effect can be parametrized as \cite{Lue:1998mq,Feng:2004mq}
\begin{eqnarray}
C_{\ell}^{TE,obs}  &=&  C_{\ell}^{TE} \cos (2 \alpha) \, ,
\label{TEobs} \\
 C_{\ell}^{TB,obs}  &=&  C_{\ell}^{TE} \sin (2 \alpha) \, ,
\label{TBobs} \\
 C_{\ell}^{EE, obs} &=& C_{\ell}^{EE}  \cos^2 (2 \alpha) +  C_{\ell}^{BB} \sin^2 (2 \alpha) \, ,
\label{EEobs} \\
 C_{\ell}^{BB, obs}  &=&  C_{\ell}^{BB}  \cos^2 (2 \alpha) +  C_{\ell}^{EE} \sin^2 (2 \alpha) \, ,
\label{BBobs} \\
 C_{\ell}^{EB, obs} &=& {1 \over 2} \left(  C_{\ell}^{EE}  - C_{\ell}^{BB}  \right) \sin (4 \alpha) \, ,
\label{EBobs}
\end{eqnarray}
with $C_{\ell}^{XY,obs} $ and $C_{\ell}^{XY}$ being the observed and the primordial (i.e. unrotated) APS for the $XY$ spectrum ($X$, $Y$ = $T$, $E$ or $B$),
i.e. the one that would arise in absence of birefringence. Note that we set the primordial $TB$ and $EB$ spectra to zero for the sake of simplicity, thereby assuming
parity violation effect plays a negligible role up to CMB photon decoupling (this choice excludes e.g. chiral gravity theories).

\begin{figure}
\centering
\includegraphics[width=0.7\hsize]{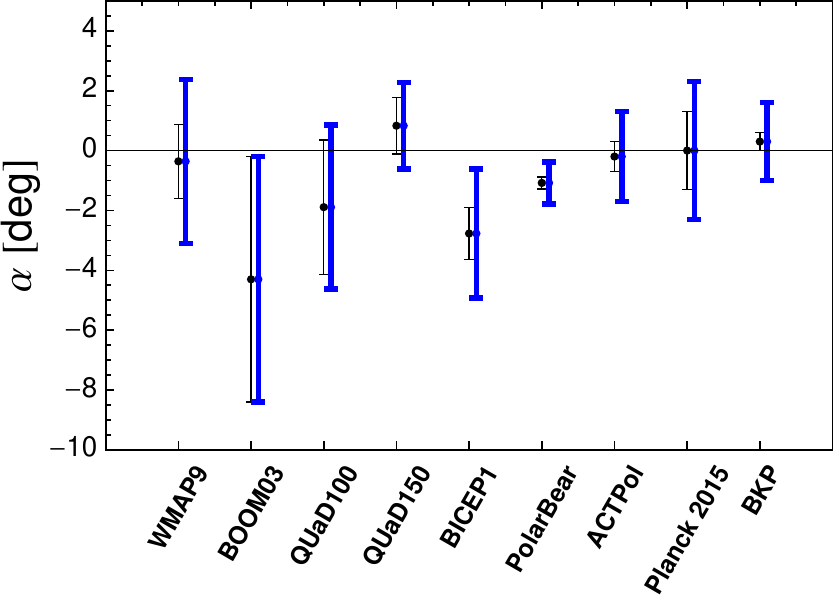}
\caption{Constraints on $\alpha$ from several CMB experiments \cite{Hinshaw:2013,Wu:2008qb,Kaufman:2013vbd,Pagano:2009kj,Ade:2014afa,Naess:2014wtr}. 
Thin black error bars are for statistical uncertainties at $68\%$ C.L.. Thick blue error bars are obtained by summing linearly the statistical and systematic uncertainties. 
The statistical uncertainty of BOOM03 do already contain a contribution from systematic effect. The last two constraints are derived in this paper.}
\label{alpha_whisker}
\end{figure}

The existing limits on $\alpha$, as obtained from CMB experiments \cite{Hinshaw:2013,Wu:2008qb,Kaufman:2013vbd,Pagano:2009kj,Ade:2014afa,Naess:2014wtr}, are shown in Fig. \ref{alpha_whisker} 
(see first seven estimates).  In short, almost all constraints are well compatible with a null effect, and the best available statistical errors are of order of $0.2^{\circ}$ with a systematic 
uncertainty estimated at the level of $0.5^{\circ}$. 
For comparison, still in Fig.~\ref{alpha_whisker}, we anticipate the constraints we obtain from {\sc Planck} and {Bicep2/Keck/Planck} (henceforth BKP) data in this paper (see the last two constraints).
{\it As discussed below the {\sc Planck} constraints do not include B information at $\ell > 29$ while this is included in BKP combination.}

This paper is organized as follows: in Section \ref{description} we provide a description of the performed analysis; 
in Section \ref{data} we report the datasets employed and the associated likelihood functions; 
in Section \ref{results} we present the main constraints on $\alpha$ sampling on all  standard $\Lambda$CDM parameters
and as well  on the tensor-to-scalar ratio, $r$. We also discuss the role of systematic uncertainties. 
We finally draw our conclusions in Section \ref{conclusion} providing also forecast for the statistical sensitivity that {\sc Planck} data can reach when BB, TB and EB spectra will be 
included at high $\ell$ in the {\sc Planck} likelihood.

\section{Description of the analysis}
\label{description}

CMB polarization arises at two distinct cosmological times: the recombination epoch ($z \simeq 1100$) and the reionization 
era at $z \sim 10$ \cite{Dodelson:2003ft}. When the CMB field is expanded in
spherical harmonics, the first signal mostly shows up at high multipoles, since polarization
is generated through a causal process and the Hubble horizon at last scattering only
subtends a degree sized angle. The later reionization of the cosmic fluid at lower redshift
impacts instead the lowest multipoles. These two regimes need to be properly treated when probing
for cosmological birefringence. 
It is also interesting to separate the constraints on $\alpha$ between high and low $\ell$ as these can be ascribed to different epochs and, hence, physical conditions.
%
Therefore, following \cite{Komatsu:2008hk}, we perform a Monte Carlo Markov chain (MCMC) analysis making use of the publicly available code \texttt{cosmomc} \cite{Lewis:2013hha, Lewis:2002ah} 
to constrain $\alpha$ by considering two multipole ranges\footnote{The transition multipole between low and high $\ell$ is taken to be $29$ following the low and high transition of the {\sc Planck} likelihood (see also Section \ref{data}).
The fine choice of this multipole does not impact significantly the results as long as it occurs in a region where the polarization signal is noise dominated \cite{Aghanim:2015xee}.}, $2 \le \ell \le 29$ 
and $ 30 \le \ell \le 2500$. 
In practice, we first sample at low $\ell$ on $\alpha$, on the optical depth $\tau$ and on the scalar amplitude $A_s$, while fixing the remaining 
$\Lambda$CDM cosmological parameters to the {\sc Planck} best-fit model derived from temperature and polarization (TT, TE, EE).
We then focus on the high $\ell$, fixing $\tau$ to the best-fit value obtained in the former analysis 
and let the remaining parameters to vary.
We call the above combination case A.  We report separately the constraints on $\alpha$ obtained by a single run on all multipoles. 
This is called case B in what follows and it is provided for comparison purposes. 
Moreover, we consider a third case, C, where we sample at the same time on two separate angles, namely $\alpha_{low}$ and $\alpha_{high}$, 
which rotate the APS independently in the two multipole regimes.

These three cases above are performed considering only scalar perturbations (i.e.~$\Lambda$CDM+$\alpha$) and 
also considering tensor perturbations (i.e.~$\Lambda$CDM+r+$\alpha$). See Table \ref{cases} for a summary of all cases considered.

Gravitational lensing induces non zero TB and EB correlations through E to B mixing, an effect potentially observable at the sensitivity level of the {\sc Planck} and BKP dataset.  
To disentangle it from cosmological birefringence, one should in principle model the latter within the CMB time evolution from last scattering to us. 
However, it can be shown \cite{Gubitosi:2014cua} that for the model given in Eqs.~(\ref{TEobs}-\ref{EBobs}), one can safely 
separate the birefringence and lensing contributions. 

\begin{table}
\centering
\caption{Summary of the considered cases. See text.}
\label{cases}
\begin{tabular}{cll}
\\
\hline
case & description & color code \\
\hline
 A & First run at low $\ell$ on $\alpha$, $A_s$ and $\tau$ & light blue\\ 
             & (other parameters fixed to Planck best fit) & \\
             & Second run at high $\ell$ on $\Lambda$CDM(+r)+$\alpha$ & blue \\ 
             & with $\tau$ fixed to the first run best fit value & \\
 B & Single run on $\Lambda$CDM(+r)+$\alpha$ in all the $\ell$ & red \\
             & range & \\
 C & Single run on $\Lambda$CDM(+r)+$\alpha_{low}$+$\alpha_{high}$  & green \\
             & where $\alpha_{low}$ (dashed) and $\alpha_{high}$ (solid) are  & \\
             & defined in the low and high $\ell$ range & \\
             & respectively & \\
\hline
\hline
\end{tabular}
\end{table}

\section{Datasets and likelihoods}
\label{data}

We have used the {\sc Planck} and BKP publicly available likelihoods\footnote{For {\sc Planck} see http://pla.esac.esa.int/pla/ and for BKP see http://bicepkeck.org/bkp\_2015\_release.html}. 
Here we list the several terms entering the likelihood functions \cite{Aghanim:2015xee,Ade:2015tva}.
\begin{itemize}
\item  \texttt{Planck lowl}. {\sc Planck} low $\ell$ temperature likelihood. Based on the {\sc Planck} 2015 low resolution CMB anisotropy map (Commander, see \cite{Adam:2015tpy}).
\item  \texttt{Planck lowTEB}. {\sc Planck} low $\ell$ temperature and polarization likelihood. Based on temperature (Commander) and polarization (70GHz) maps.
\item  \texttt{Planck TT,TE,EE}. {\sc Planck} high $\ell$ temperature and polarization Likelihood. Based on TT, TE and EE cross-spectra.
\item \texttt{BKP BB}. Bicep2/Keck likelihood with 353GHz {\sc Planck} channel. Based on the first five band powers of the BB spectra \cite{Ade:2015tva}.
\item \texttt{BKP EE,BB}. Bicep2/Keck likelihood with 353GHz {\sc Planck} channel. Based on the first five band powers of the BB, EE and EB spectra.
\end{itemize}

From Eqs.~(\ref{TEobs}-\ref{EBobs}) it is clear that we are sensitive to the sign of $\alpha$ only when we use TB and EB information. 
This happens only when we employ \texttt{Planck lowTEB} and \texttt{BKP EE,BB}. Therefore, when the latter are not considered, in fact our constraints on $\alpha$ will be on its absolute value.

For each considered case, we employ the following likelihoods:
\begin{itemize}
\item Case A. First run \texttt{Planck lowTEB}. Second run \texttt{Planck lowl} + \texttt{Planck TT,TE,EE}. 
\item Case B. Single run with \texttt{Planck lowTEB} +  \texttt{Planck TT,TE,EE}.
\item Case C. Single run with \texttt{Planck lowTEB} +  \texttt{Planck TT,TE,EE}.
\end{itemize}
All the above cases are analyzed for both $\Lambda$CDM+$\alpha$ and $\Lambda$CDM+r+$\alpha$.
Moreover, in the latter case we consider the likelihoods \texttt{BKP BB} or \texttt{BKP EE,BB} either alone (with minimum multipole $\ell \simeq 50$) or in combination with {\sc Planck}.


\section{Results}
\label{results}

\subsection{$\Lambda$CDM+$\alpha$}
\label{lcdmalpha}

The $\Lambda$CDM+$\alpha$ model is analyzed considering {\sc Planck} data alone.
\begin{figure}
\centering
\includegraphics[width=0.45\hsize]{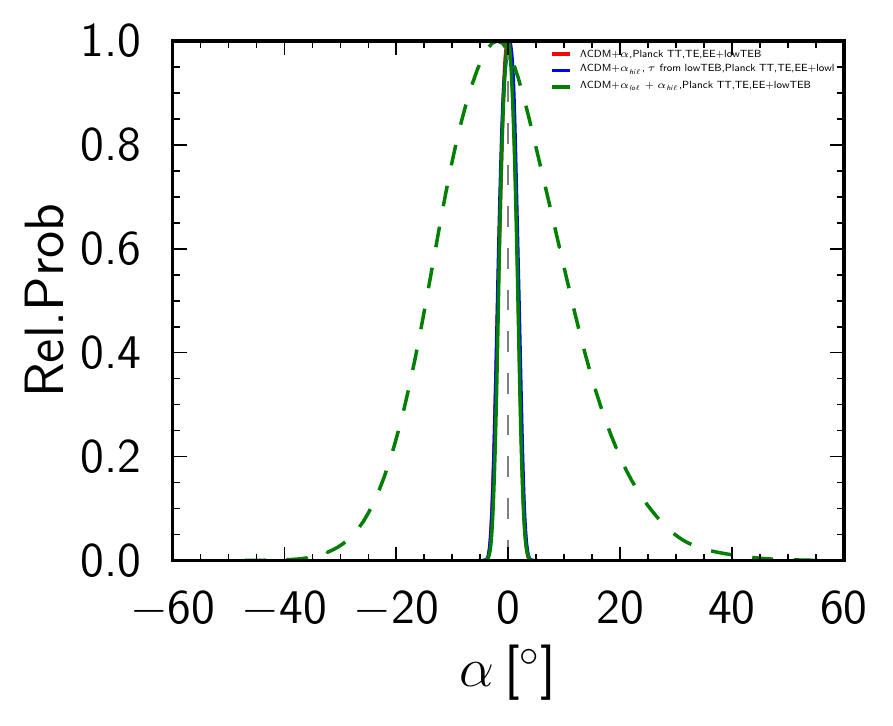}
\includegraphics[width=0.45\hsize]{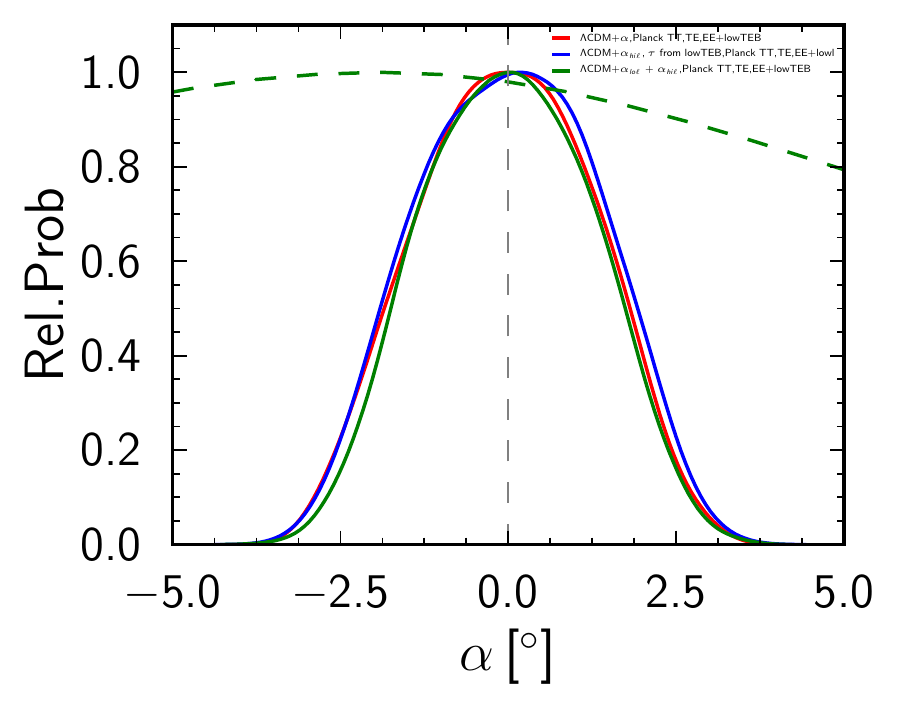}
\caption{Left panel: Constraints on $\alpha$ from the three considered cases summarized in Table \ref{cases}. Blue curves are for case A, red curves for case B
and green curves for case C (solid green for $\ell \in [30,2500]$ and dashed green for $\ell \in [2,29]$). Right panel: zoom of the left panel.}
\label{fig:biri}
\end{figure}
\begin{figure}
\centering
\includegraphics[width=\hsize]{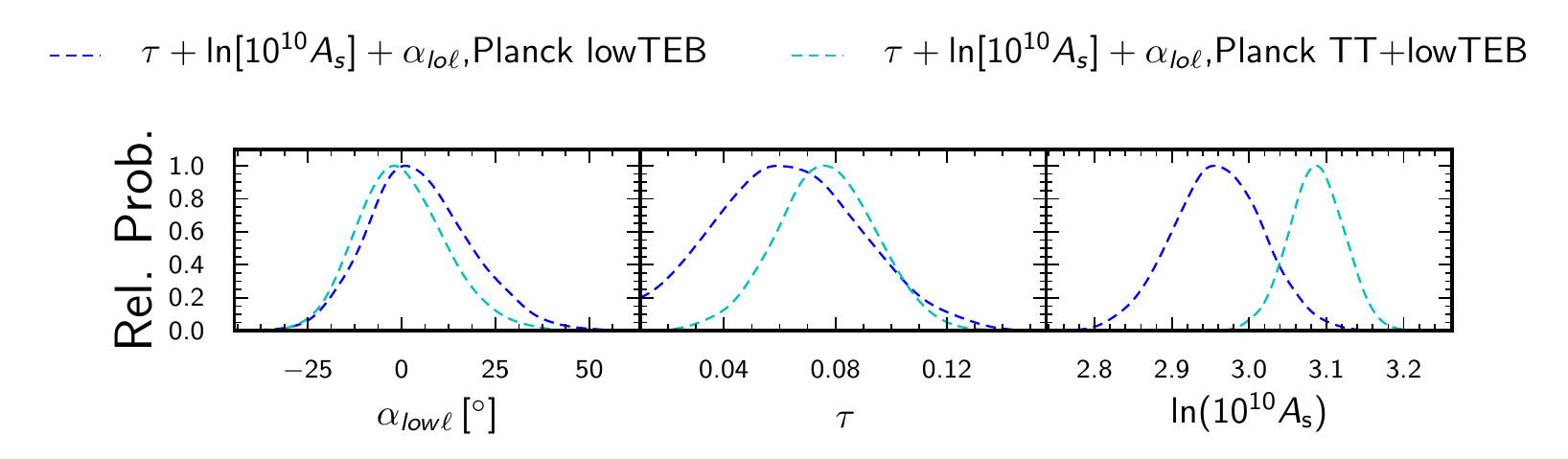}
\caption{Posterior distributions of $\alpha$, $\tau$ and $A_s$ for the low $\ell$ run of case A. 
Blue curves consider temperature and polarization at low $\ell$ while light blue take into account polarization at low $\ell$ and temperature at all $\ell$.
}
\label{fig:casoalowell}
\end{figure}
The posterior distributions for $\alpha$ are shown in Fig.~\ref{fig:biri} and the corresponding constraints are reported in Table \ref{tabellauno}.
The latter only account for statistical uncertainty ($68\%$ C.L.). 
The companion ``low-$\ell$ run'' of case A is displayed in Fig.~\ref{fig:casoalowell}.
\begin{table}
\footnotesize
\centering
\caption{Constraints of $\alpha$ for the posterior distributions shown in Fig.~\ref{fig:biri}. Units: deg.
{\sc P} stands for {\sc Planck}, BKP EE and BKP EE BB for Bicep2/Keck/Planck BB and EE,BB,EB data.
unc = unconstrained. $\star$ corresponds to the case in which the six $\Lambda$CDM parameters are fixed to the {\sc Planck} best fit.}
\label{tabellauno}
\begin{tabular}{cccccc}
\\
\hline
\hline
case & $\alpha_{low}$ & $\alpha_{high}$ & $\alpha$ & model & data \\
          &                             &                               &                  & $\Lambda$CDM & \\
\hline
\hline
 A & $5^{+12}_{-15}$ & $0.0 \pm 1.4$ & - & +$\alpha$ & {\sc P}\\
 B & - & - & $-0.03 \pm 1.35$ & +$\alpha$ & {\sc P}\\
 C & $-0.5^{+10.2}_{-12.7}$ & $0.0 \pm 1.3$ & -& +$\alpha$ & {\sc P} \\
 A & $4^{+12}_{-15}$ & $0.0 \pm 1.4$ & - & +$\alpha$+r & {\sc P}\\
 B & - & - & $0.00 \pm 1.33$ & +$\alpha$+r & {\sc P}\\
 C & $0.6^{+9.8}_{-13.6}$ & $0.0 \pm 1.3$ & -& +$\alpha$+r & {\sc P} \\
 A & $4^{+12}_{-15}$ & $0.0 \pm 1.4$ & - & +$\alpha$+r & {\sc P}+BKP BB\\
 B & - & - & $0.02^{+1.31}_{-1.41}$ & +$\alpha$+r & {\sc P} + BKP BB\\
 C & $0.5^{+9.8}_{-10.0}$ & $0.0 \pm 1.4$ & -& +$\alpha$+r & {\sc P} +BKP BB\\
 A & $4^{+12}_{-15}$ & $0.32 \pm 0.27$ & - & +$\alpha$+r & {\sc P}+BKP BB EE\\
 B & - & - & $0.32 \pm 0.26$ & +$\alpha$+r & {\sc P} + BKP BB EE\\
 C & $0.9^{+9.9}_{-13.6}$ & $0.32 \pm 0.26$ & -& +$\alpha$+r & {\sc P} +BKP BB EE\\
 \, C$^{\star}$ & unc & $0.0 \pm 1.5$ & -& +$\alpha$+r & BKP BB\\
 \, C$^{\star}$ & unc & $0.30 \pm 0.27$ & -& +$\alpha$+r & BKP BB EE\\
\hline
\hline
\end{tabular}
\end{table}
\begin{figure}
\centering
\includegraphics[width=\hsize]{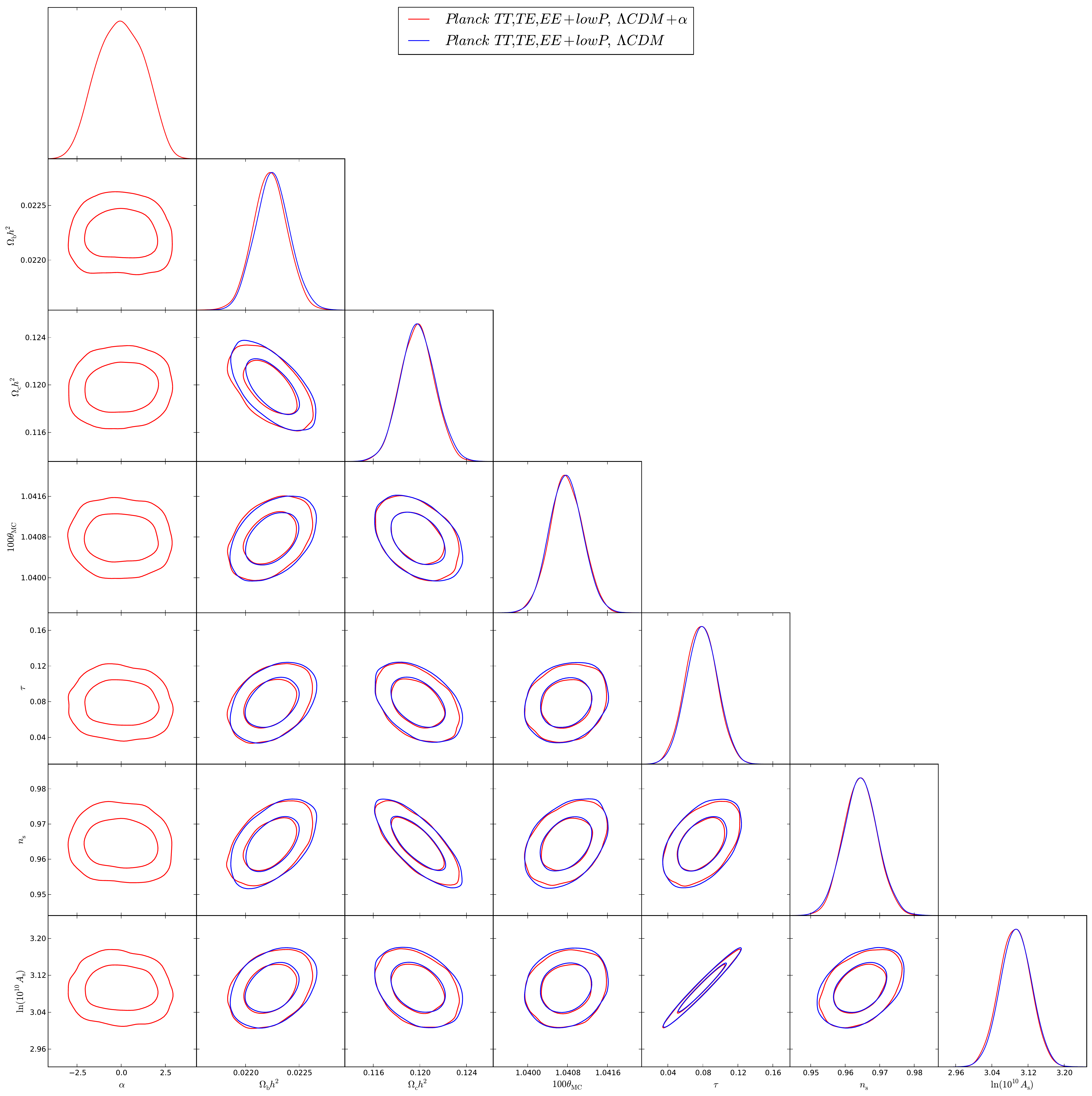}
\caption{Comparison of the base $\Lambda$CDM model parameters and the birefringence angle $\alpha$ from the analysis over the $2 \le \ell < 2500$ range (case B).
All the 2D contour plots are very stable against the inclusion of $\alpha$ in the MCMC sampling. We conclude that $\alpha$ is largely uncorrelated with the other $\Lambda$CDM parameters.
\label{fig:tri}
}
\end{figure}
\begin{figure}
\centering
\includegraphics[width=\hsize]{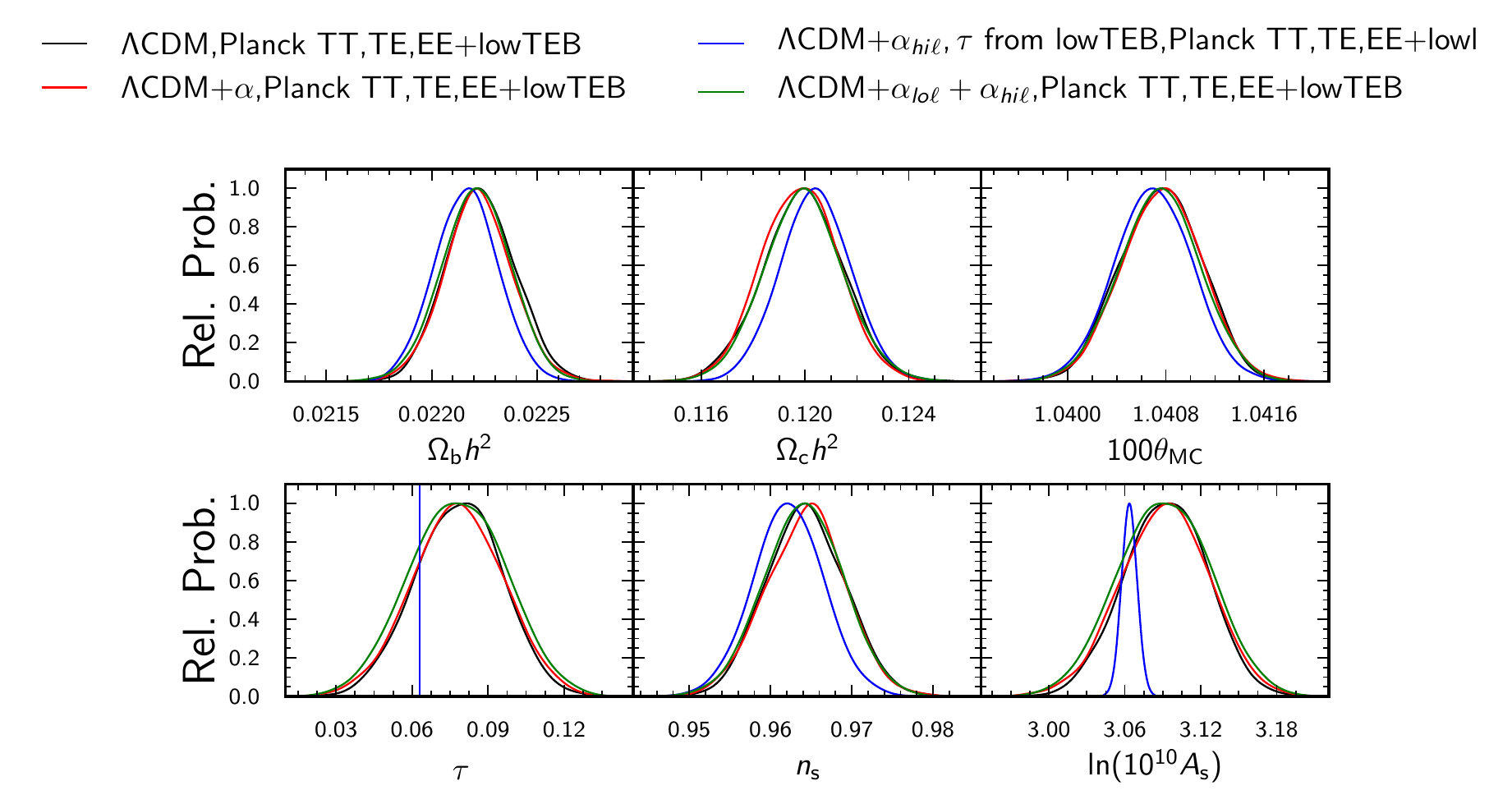}
\caption{Marginalized posterior distributions for the six main cosmological parameters for the $\Lambda {\rm CDM}$ when jointly sampled with $\alpha$ for the three considered cases. 
All the parameters are very compatible among the three cases.
\label{fig:cosmo}
}
\end{figure}
Fig.~\ref{fig:tri} shows that all the standard $\Lambda {\rm CDM}$ cosmological parameters are very stable against the inclusion of $\alpha$ in the MCMC
and Fig.~\ref{fig:cosmo} shows how the same parameters are very stable in all three considered cases.

All the {\sc Planck} constraints are compatible with zero even considering only the statistical uncertainty.
Still without accounting for systematic uncertainty, the best constraint for {\sc Planck} alone is 
$\alpha = 0.0^{\circ} \pm 1.3^{\circ}$. 
The standard error will improve significantly in a future version of the high $\ell$ likelihood which will employ the TB and EB spectra. Forecasts are provided below.

\subsection{$\Lambda$CDM+r+$\alpha$}
\label{lcdmralpha}
The $\Lambda$CDM+r+$\alpha$ model is analyzed with the following data set combinations: {\sc Planck} alone,  BKP and {\sc Planck} plus BKP. Since all data in the BKP dataset come from the small 
($f_{sky} \simeq 1 \%$) region observed by Bicep, we neglect the bias caused by counting twice the {\sc Planck} dataset in this region, for the latter combination.
As for the $\Lambda$CDM+$\alpha$ model, all the estimates are reported in Table \ref{tabellauno}.
\begin{itemize}
\item {\sc Planck} data alone. 
The posterior distribution functions of $\alpha$ are displayed in Fig.~\ref{fig:biri+r}.
\begin{figure}
\centering
\includegraphics[width=0.45\hsize]{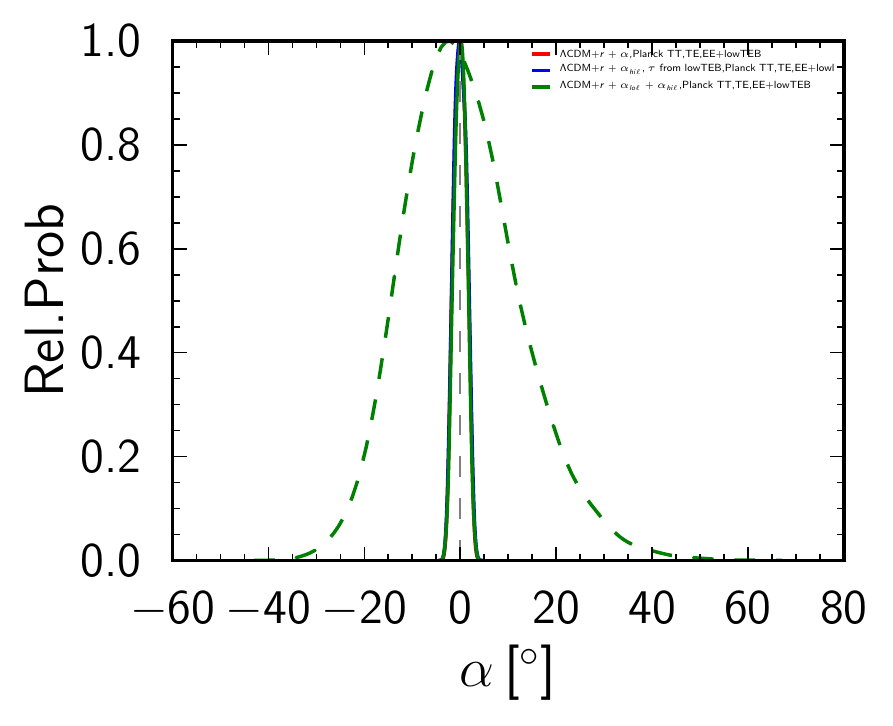}
\includegraphics[width=0.45\hsize]{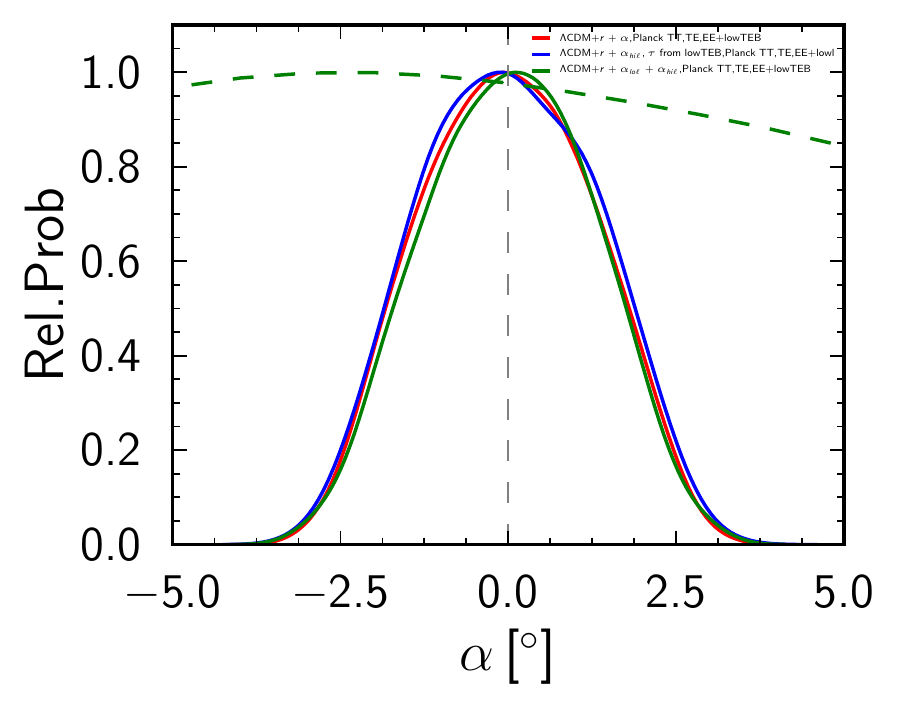}
\caption{Left panel: Constraints on $\alpha$ from the three considered cases taken into account in $\Lambda$CDM+r+$\alpha$. Blue curves are for case A, red curves for case B 
and green curves for case C (solid green for $30-2500$ and dashed green for $2-29$). Right panel is just a zoom of the right panel. Only {\sc Planck} data are employed.}
\label{fig:biri+r}
\end{figure}
In Fig.~\ref{fig:casoalowellpr} we show the low $\ell$ companion run of case A.
\begin{figure}
\centering
\includegraphics[width=0.9\hsize]{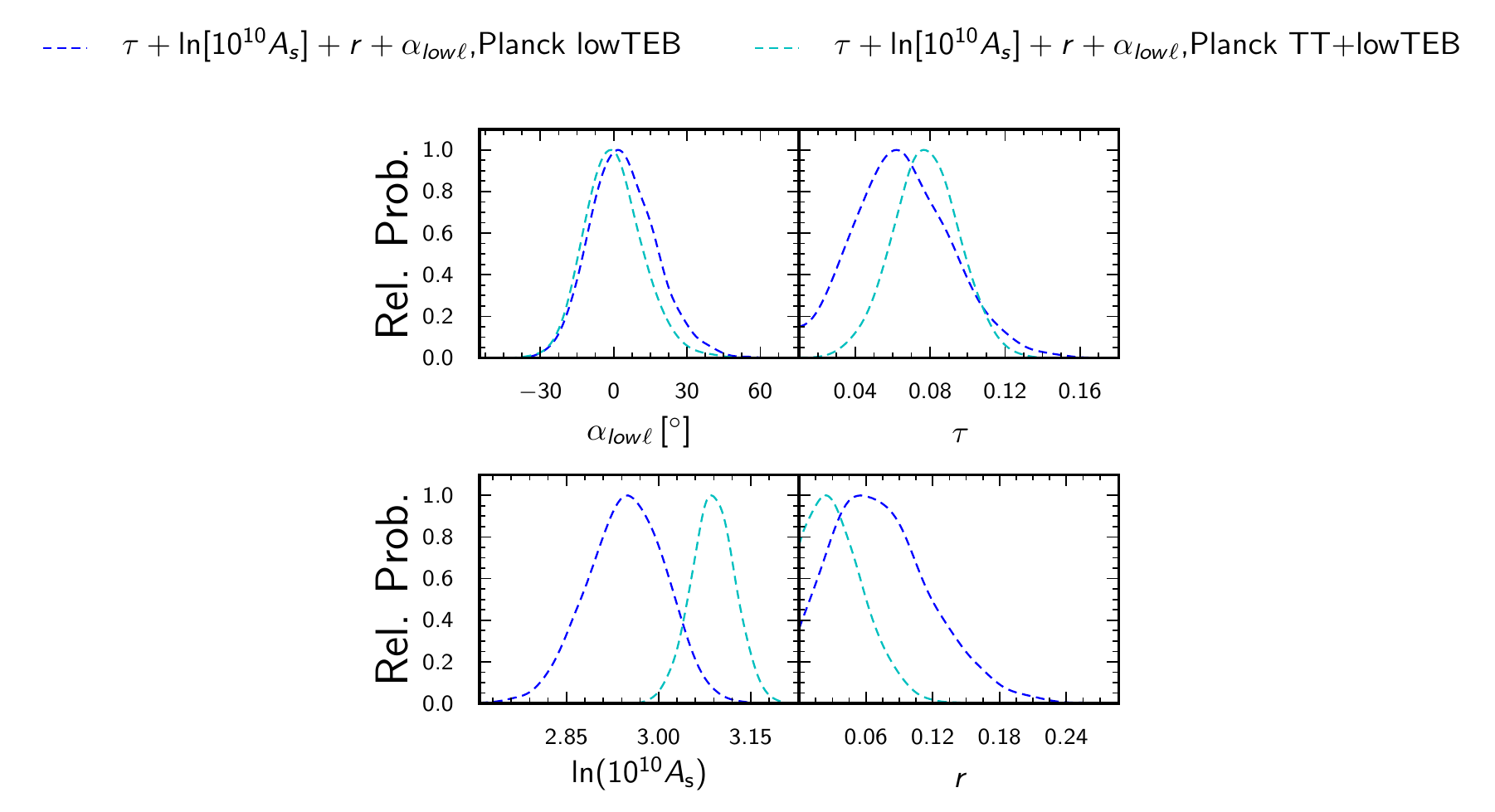}
\caption{Low $\ell$ run of case A. In $\Lambda$CDM+r+$\alpha$.}
\label{fig:casoalowellpr}
\end{figure}
The posterior distribution functions of the six standard cosmological parameters and $r$ are plotted in Fig.~\ref{fig:cosmo+r}.
All the cases are very stable and consistent. The inclusion of $\alpha$ does not produce any shift in the $\Lambda$CDM+$r$ parameters. 
\begin{figure}
\centering
\includegraphics[width=\hsize]{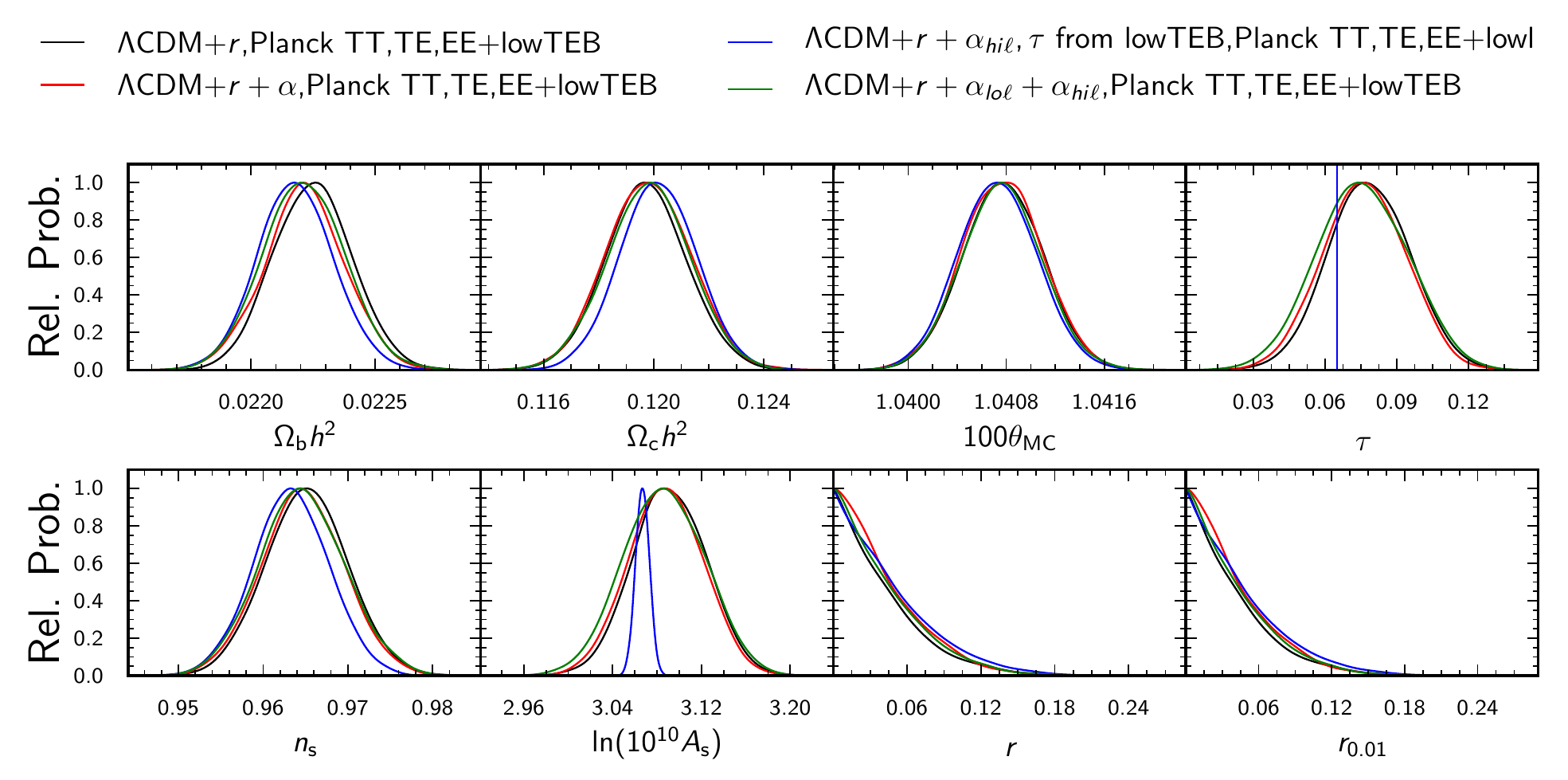}
\caption{Marginalized posterior distributions for the seven main cosmological parameters for the $\Lambda {\rm CDM}$+r (blue) and $\Lambda {\rm CDM}+r+\alpha$ (red) models from the analysis over the $2 \le \ell < 2500$ range. It is clear that the parameters are very stable against the inclusion of $\alpha$.
\label{fig:cosmo+r}
}
\end{figure}
As for the $\Lambda$CDM+$\alpha$ model, even in the $\Lambda$CDM+r+$\alpha$ model all the {\sc Planck} constraints 
on $\alpha$ are compatible with zero and numerically very similar.
\item {\sc Planck} and BKP data. The probability distribution functions of $\alpha$ are shown in Fig.~\ref{fig:biri+r_planckbkp_bkp}.
Blue, red and green solid (dotted-dashed) curves are for {\sc Planck} + BKP BB (BKP EE BB) data in the cases A, B and C respectively. 
Cyan curves are for BKP data alone (solid for BKP BB and dotted-dashed for BKP EE BB).
The corresponding constraints are reported in Table \ref{tabellauno}.
When {\sc Planck} data are employed in combination with BKP BB we notice a change in the shape of the probability distribution function and a bimodal distribution appears. This is due to the fact that the sign of $\alpha_{high}$ is not constrained when EB and TB spectra are neglected.
However the compatibility with zero is at the same level as in the {\sc Planck} data alone analysis.
When BKP EE BB data are taken into account in combination with {\sc Planck}, due to the presence of the EB spectrum, the bimodal pattern disappears and a better constraint of $\alpha$ is obtained, i.e. $\alpha = 0.32^{\circ} \pm 0.26^{\circ}$ (again, only statistical). In this case the BKP data dominates the combination and the impact of {\sc Planck} data is subleading
since what counts more is the EB spectrum.
\begin{figure}
\centering
\includegraphics[width=0.8\hsize]{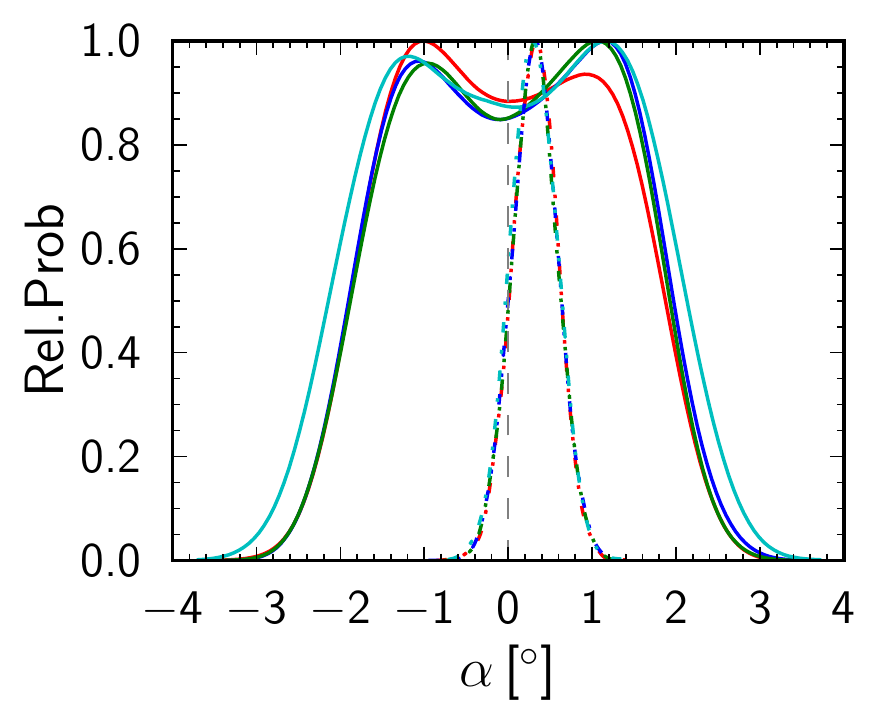}
\caption{Probability distribution function for $\alpha$. Blue, red and green solid (dotted-dashed) curves are for {\sc Planck} + BKP BB (BKP EE BB) data in the cases A, B and C respectively. 
Cyan curves are for BKP data alone (solid for BKP BB and dotted-dashed for BKP EE BB).}
\label{fig:biri+r_planckbkp_bkp}
\end{figure}
\end{itemize}

\subsection{Uncertainties from systematic effects}
\label{sys}
The most relevant untreated systematic effect that impacts birefringence constraints is a miscalibration of the detectors' optimal polarization angle, i.e. the angle, measured in the telescope reference frame, at which a detector is maximally sensitive to a vertical linear polarization \cite{Pagano:2009kj}. In the case of {\sc Planck}, ground based measurements cannot be assumed to be representative of in-flight conditions at face value, but rather need to be extrapolated since thermoelastic effects arising in operations can play a role. In-flight measurements are complicated by the scarsity of linearly polarized sources that are bright enough, though the Crab nebula has been attempted as a calibrator \cite{Adam:2015vua}. 
For  {\sc Planck}, we use the estimate of the error in the polarization orientation evaluated in \cite{Rosset:2010vc}, where it is given at $1^{\circ}$.
The original Bicep2 analysis \cite{Ade:2014xna} discusses how the TB and EB spectra are rotated to minimize E to B mode mixing, by imposing that their residuals from zero are minimized. This process destroys of course any detection power for birefringence below this final rotation angle, which is estimated at $1^{\circ}$ and can thus be taken as a measure of the final systematic budget. In the BKP analysis, the treatment of the Bicep and Keck power spectra is identical to that of Bicep2. We therefore assume a $1^{\circ}$ systematic uncertainty of the instrumental polarization angle for the BKP dataset as well\footnote{Notice that for Keck the rotation angle is smaller, i.e.\ about $0.5^{\circ}$ degrees \cite{Ade:2015fwj}. We assume therefore that the error is dominated by Bicep.}. 
The same assumption applies to the joint analysis of BKP and {\sc Planck}.

\section{Conclusions}
\label{conclusion}

We have constrained the birefringence angle $\alpha$ employing {\sc Planck} and BKP data either alone or in combination. 
All our results are well compatible with no detection even without accounting for systematic uncertainty.
We find $\alpha = 0.0^{\circ} \pm 1.3^{\circ} \mbox{ (stat)} \pm 1^{\circ}  \mbox{ (sys)} $ for {\sc Planck} data alone. 
Similar results are found when we also consider the BB spectrum of BKP data. 
The inclusion of the EB spectrum in the BKP likelihood makes the constraint on $\alpha$ much tighter, namely $\alpha = 0.32^{\circ} \pm 0.26^{\circ} \mbox{(stat.)} \pm 1^{\circ} \mbox{(sys.)}$.
In the latter case the inclusion of {\sc Planck} data brings little constraining power. 
This is expected since the current version of the {\sc Planck} Likelihood does not contain TB and EB spectra at high $\ell$.
It is possible to forecast that including BB, TB and EB spectra in the high $\ell$ {\sc Planck} likelihood and considering the range of multipoles, i.e. [30-2500], 
the level of the statistical uncertainty on $\alpha$ can be as low as $\sim 0.03^{\circ}$. This is obtained considering the nominal sensitivity of the $143$ GHz channel 
and an exact likelihood approach, see e.g. \cite{Perotto:2006rj}. The obtained value is compatible with what forecasted in the {\sc Planck} bluebook \cite{Planck:2006aa}
considering the longer integration time and the improved sensitivity of the used channel. 
Of course, the quest for reducing the statistical errors would eventually prove meaningless if it is not accompanied by a parallel effort to understand and reduce the systematic error budget affecting polarization rotation constraints.

\section{Acknowledgements}
This paper is based on observations obtained with the satellite {\sc Planck} (http://www.esa.int/Planck), an ESA science mission with instruments and contributions directly funded by ESA Member States, NASA, and Canada.
We acknowledge the use of computing facilities at NERSC (USA), of the HEALPix package [23], and of the Planck Legacy Archive (PLA). 
Research supported by ASI through ASI/INAF Agreement I/072/09/0 for the Planck LFI Activity of Phase E2.
MG and LP acknowledges support by the research grant Theoretical Astroparticle Physics number 2012CPPYP7 under the program PRIN 2012 funded by MIUR and by TASP (I.S. INFN).

\end{document}